\documentclass[12pt,preprint,english]{aastex}
\usepackage[T1]{fontenc}
\usepackage[latin1]{inputenc}
\usepackage{graphicx}

\makeatletter

\providecommand{\tabularnewline}{\\}

\usepackage{times}

\usepackage{babel}
\makeatother
\begin{document}

\title{The Sensitivity of Hybrid Differential Stereoscopy for Spectral Imaging}

\author{C.E. DeForest and C.C. Kankelborg}

\begin{abstract}
Stereoscopic spectral imaging is an observing technique that affords
rapid acquisition of limited spectral information over an entire image
plane simultaneously. Light from a telescope is dispersed into multiple
spectral orders, which are imaged separately, and two or more of the
dispersed images are combined using an analogy between the $(x,y,\lambda)$
spectral data space and conventional $(x,y,z)$ three-space. Because
no photons are deliberately destroyed during image acquisition, the
technique is much more photon-efficient in some observing regimes
than existing techniques such as scanned-filtergraph or scanned-slit
spectral imaging. Hybrid differential stereoscopy, which uses a combination
of conventional cross-correlation stereoscopy and linear approximation
theory to extract the central wavelength of a spectral line, has been
used to produce solar Stokes-V (line-of-sight) magnetograms in the
617.34 nm Fe I line, and more sophisticated inversion techniques are
currently being used to derive Doppler and line separation data from
EUV images of the solar corona collected in the neighboring lines
of He-II and Si-XI at 30.4 nm. In this paper we develop an analytic
\emph{a priori} treatment of noise in the line shift signal derived
from hybrid differential stereoscopy. We use the analysis to estimate
the noise level and measurement precision in a high resolution solar
magnetograph based on stereoscopic spectral imaging, compare those
estimates to a test observation made in 2003, and discuss implications
for future instruments.
\end{abstract}


\section{Introduction}

Spectral imaging in general, and solar spectral imaging in particular,
suffer from a fundamental problem in detector physics. Spectral images
have three independent variables $(x,y,\lambda)$, while current image
detectors only support two independent variables $(x,y)$ and integrate
over wavelength $\lambda$. Conventional techniques to overcome this
problem generally use time-multiplexing: in filtergraph imaging spectroscopy,
a narrow band filter is tuned slowly across the spectral range of
interest and an image collected at each discrete $\lambda$; in conventional
scanned-slit imaging spectroscopy, the light is passed through a spatial
filter (the slit), selecting a single $x$, and the remaining light
is dispersed to project $\lambda$ onto the detector's $x$ axis.
Even more sophisticated techniques such as Fourier imaging spectroscopy
use time multiplexing to collect multiple two-dimensional basis images
of the three-dimensional data space. Multiplexing in time is photon-inefficient
as photons that are not part of the current sample are discarded.
That is a problem because solar measurements are photon starved: instruments
must race to collect sufficient photons for a spectral measurement,
before the features on the Sun change. 

Stereoscopic spectral imaging overcomes the problem of spectral imaging
by analogy between the spectral imaging problem and the stereoscopic
problem of determining feature distance in ordinary 3-space. Dispersed
spectral images are integrals along diagonal lines in $(x,y,\lambda)$
space; they are analogous to images of a 3-space subject, with the
images collected at a {}``look angle'' that depends on the dispersion
of the instrument. Collecting multiple dispersed images in different
spectral orders yields data that can be inverted stereoscopically
to measure some spectral characteristics of the line everywhere in
the image plane simultaneously.

\citet{kankelborg2001} and \citet{Fox2003} have described using
this stereoscopic approach to simultaneous imaging and spectroscopy
in EUV emission lines. Their \emph{MOSES} rocket payload, launched
2006 February 8, obtained Doppler measurements of explosive events
and jets in the He II line at $30.4$ nm (\citealt{Kankelborg2007}).
Another related technique, computed-tomography imaging spectroscopy
(CTIS) uses multiple spectral orders for hyperspectral imaging and
spectropolarimetry at low spectral resolution in the visible and infrared
(e.g. \citealt{Wilson1997,Miles1999,Milesetal1999,Dereniak2005}). 

DeForest et al. (2004) have developed a first-order theory of stereoscopic
inversion that is applicable to a much simpler spectral context: measuring
Doppler shift and the longitudinal Zeeman splitting in a narrow visible
absorption line in the solar photosphere. That work is conceptually
similar to the \emph{MOSES} stereoscopic imaging, but there are two
key differences: there is only a single spectral line in the instrument
passband, and it is an absorption line; and the line width is narrower
than a pixel in dispersion. DeForest et al. demonstrated the technique
with proof-of-concept observations of the quiet Sun, a decayed active
region, and a clean sunspot. However, in that work they did not analyze
the noise characteristics of the observing and inversion technique,
only report the \emph{a posteriori} noise measured in a quiet Sun
region. 

In this article, we derive expressions for the \emph{a priori} noise
level (hence sensitivity) and systematic error of a stereoscopic instrument
in the narrow-line regime described by \citet{DeForestetal2004} using
hybrid differential stereoscopy to determine line shift. We use \emph{a
priori} statistical calculation to derive the effect of the dominant
source of uncorrelated noise - photon counting statistics - and also
address quasi-random systematic errors in the inversion using some
basic assumptions about the scene being viewed. 

The analytic expressions are general and may be applied to
stereoscopic spectrographs viewing either absorption or emission
lines, but we consider them in the specific context of an
absorption-line Zeeman magnetograph viewing the relative line shift
between the right- and left-circular polarizations. In \S
\ref{sec:Hybrid-differential-stereoscopy} we analyze each of several
noise sources individually; in \S \ref{sec:Summary-&-Specific} the
results of the individual analyses are applied to generate a noise
budget for an example baseline instrument observing the quiet
photosphere with \textasciitilde{}0.1 arc sec resolution and 0.03 arc
sec pixels,in the 617.34 nm Fe I absorption line. Finally, in \S
\ref{sec:Conclusions} we draw conclusions about the types of
observation for which hybrid differential stereoscopic spectral
imaging offers the best prospects for advances over the state of the
art.  Throughout the discussion we have used $\Delta$ to indicate a
single difference from the correct or expected value of the following
quantity, and angle-brackets to denote RMS averages over an image.

\section{\label{sec:Hybrid-differential-stereoscopy}Hybrid differential stereoscopy}

Here we develop a theory of noise in a hybrid differential analysis
of absorption line data. Following \citet{DeForestetal2004}, consider
an image in the $(x,y)$ plane produced by observing, through a slitless
dispersing spectrograph, a narrow absorption line of rest wavelength
$\Lambda_{0}$, offset $\Lambda'(x,y)$ from that wavelength, constant
width $\Delta\lambda$, and integrated total intensity absorption
$L(x,y)$ over that width; limiting the wavelength range on the detector
is a filter with transmission profile $F(\lambda')$ and total admitted
continuum intensity $C(x,y)$ in the absence of the spectral line.
Without loss of generality, take the dispersion direction of the spectrograph
to be along the $x$ axis and consider the images collected in the
$\pm1$ spectral order. Then the $y$ variation of the image drops
out of the analysis and inversion may be performed independently along
each line parallel to the $x$ axis. Variations in the central wavelength
of the spectral line result in spatial distortions of the images;
these distortions are antisymmetric across the two spectral orders,
and can be used to recover low spatial frequencies of the function
$\Lambda'(x)$ by \emph{correlation stereoscopy}. Small patches of
the images are cross-correlated to determine the offset functions
$X_{L,\pm1}(x)$, and the wavelength offset can then be determined
by \begin{equation}
\Lambda'_{L}=\frac{X_{L,+1}-X_{L,-1}}{2\alpha},\label{eq:low-freq dispersion}\end{equation}
where $\alpha$ is the dispersion of the +1 order. The $L$ subscripts
refer to low spatial frequencies; the maximum spatial frequency that
may be resolved by this method is determined by the patch size used
for the cross-correlation. The inversion used for Eq. \ref{eq:low-freq dispersion}
is analogous to the inversion carried out by the human visual system,
determining the $z$ coordinate of observed objects by cross-correlating
the visual fields from the left and right eye. Recovering higher spatial
frequencies requires a more sophisticated inversion. DeForest (2003)
and DeForest et al. (2004) developed the process of \emph{differential
stereoscopy} to determine high spatial frequencies in $\Lambda'$
based on first-order intensity variation in the dispersed images.
The difference between the +1 and -1 order dispersed images yields
several terms in first approximation order, one of which is proportional
to $d\Lambda'/dx$. By integrating the difference signal along the
$x$ direction it is possible to recover an approximation to the high
spatial frequencies $\Lambda'_{H}$ throughout the image plane --
because the primary detection term is to the spatial derivative, low
spatial frequencies are not well measured with that technique. Furthermore,
differential stereoscopy is sensitive to {}``leakage'' of intensity
and other signals into the $d\Lambda'/dx$ measurement. These terms
are proportional to $\Lambda'$ itself and must therefore be compensated
in some way in strong field regions. 

By resampling the dispersed images to eliminate the spatial shifts
related to $\Lambda'_{L}$ before applying differential stereoscopy,
it is possible to eliminate image-related noise in the differential
signal, to first approximation order. The differential process is
used to recover high spatial frequencies, and the low spatial frequencies
are inserted via a mathematical filter. The resulting \emph{symmetric
hybrid inversion equation} for first-order stereoscopic inversion
(from DeForest et al. 2004) is:\begin{equation}
\Lambda'_{c1}=\int\left\{ \frac{I_{-1}\left(X_{L,-1}\right)-I_{+1}\left(X_{L,+1}\right)}{\alpha\left(I_{+1}+I_{-1}\right)}\frac{C}{L}-\epsilon\left(\Lambda'_{c1}-\Lambda'_{L}\right)\right\} dx\label{eq:hybrid_inversion}\end{equation}
where $\Lambda'_{L}$ is again the low-spatial-frequency component
of the central-wavelength shift image, as determined by explicit cross-correlation
of patches of the +1 image with corresponding regions of the -1 image,
$X_{L,-1}$, and $X_{L,+1}$ are again distorted spatial coordinates
that use the correlation signal to remove measured offsets in the
images, $I_{\pm1}$ is the intensity profile of each image as a function
of focal-plane coordinate $x$ or the distorted focal-plane coordinate
$X_{L,\pm1}$, $\alpha$ is the dispersion of the instrument, and
$\epsilon$ is a small convergence factor that splices the frequency
spectra of the two derivation methods. 

Note that Eq. \ref{eq:hybrid_inversion} is implicit rather than closed-form:
$\Lambda'_{c1}$is present both on the left hand side of the equality,
and also inside the integral on the right hand side. The presence
of the corrective filter term $\epsilon(\Lambda'_{c1}-\Lambda'_{L})$
is what makes Eq. \ref{eq:hybrid_inversion} a \emph{hybrid} inversion:
it preserves the high spatial frequencies present in the main portion
of the integrand, while forcing the low spatial frequencies in the
integral to match the $\Lambda'_{L}$ signal obtained by direct cross-correlation
of image patches. If $\Lambda'_{L}$ happens to be constant and zero,
then the impulse response function of the resulting spatial filter
is a decaying exponential. In practice, the integral should be performed
numerically in both the positive-X and negative-X direction, and the
two results summed. The forward integral is causal along the $x$
axis and the reverse integral is anti-causal, so that the resulting
impulse response function is symmetric in $x$.

The discrete form of Equation \ref{eq:hybrid_inversion}, then, is
best calculated using the sum of two discrete filter transforms. Let
$J$ be the normalized difference image in compensated coordinates,
interpolated to (integer) non-distorted pixel locations, and let $i$
run over the pixel coordinates in the $x$ direction. Then $J$ may
be written:\begin{equation}
J(i)\equiv\frac{I_{-1X}(i)-I_{+1X}(i)}{\left(I_{-1X}(i)+I_{+1X}(i)\right)}\label{eq:J-definition}\end{equation}
where the $X$ subscript is to indicate that the $I_{\pm1}$ images
have been interpolated so that the $i^{th}$ pixel is found from the
focal-plane location $x_{i}\pm\alpha\Lambda_{L}(x_{i}).$ In the first-order
theory developed by DeForest et al. 2004, $J$ is important because,
when $d\Lambda'/dx\ll\alpha^{-1},$\begin{equation}
J\sim\alpha\frac{L}{C}\frac{d\Lambda'}{dx}\label{eq:dlambda/dx}\end{equation}
which allows recovery of $\Lambda'$ by integrating and scaling appropriately.

The integral becomes a sum of two discrete spatial filters in the
\emph{discrete symmetric hybrid inversion equation}:\begin{equation}
\Lambda'_{c1}(i)=\frac{1}{2\alpha}\left(\sum_{n=0}^{i}\left(\left\langle \frac{C}{L}\right\rangle_{RMS} J(n)+\epsilon\Lambda'_{L}(n)\right)\left(1-\epsilon\right)^{i-n}-\sum_{n=i}^{w-1}\left(\left\langle \frac{C}{L}\right\rangle_{RMS} J(n)-\epsilon\Lambda_{L}^{'}(n)\right)\left(1-\epsilon\right)^{n-i}\right)\label{eq:discrete inversion}\end{equation}
where pixel centers are considered to exist at all integers between
0 and $w$, and$\left\langle C/L\right\rangle_{RMS} $ is the average ratio
of admitted continuum intensity to line absorption intensity. The
$\epsilon\Lambda_{L}'$ terms merely produce a smoothed version of
$\Lambda'_{L}$: if $\Lambda'_{L}$ were constant, and the image were
infinite in extent, then each of the two summations would return exactly
$\Lambda'_{L}$ by the law of geometric sums. The low pass filter
from this smoothing effect exactly matches the high-pass filter applied
to the integral by the exponential decay term, so that the two versions
of $\Lambda'$ are combined together seamlessly. Equation \ref{eq:discrete inversion},
together with the process of cross-correlation to determine the relationship
between $X_{L,\pm1}$ and $x$, is a complete discrete image inversion. 

\begin{figure}
\includegraphics{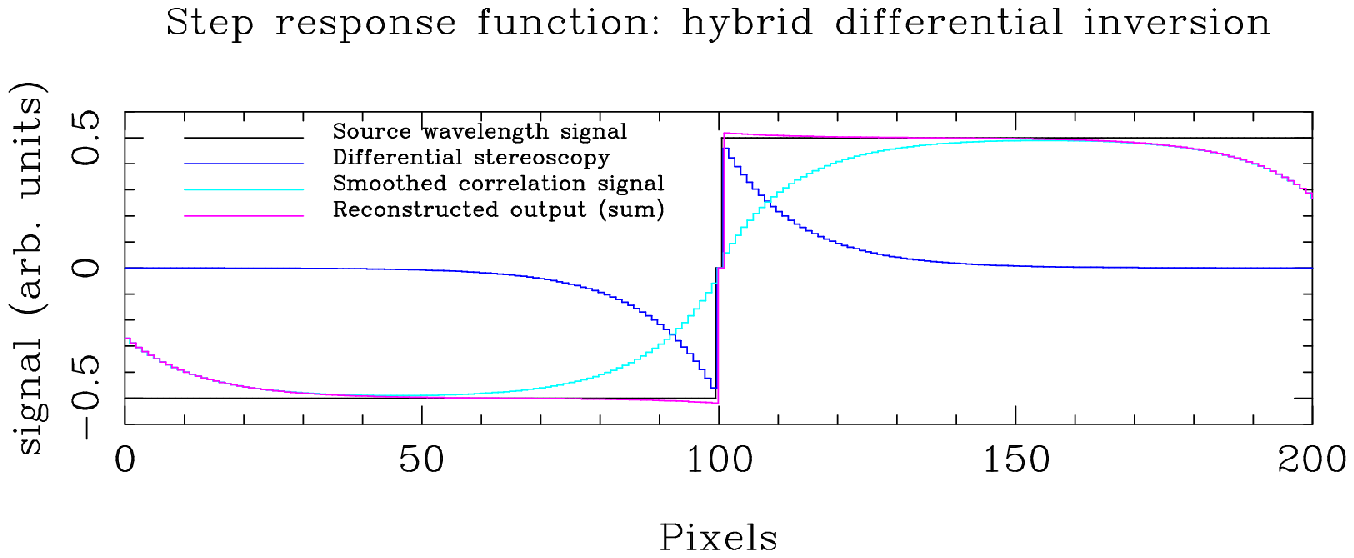}

\caption{\label{fig:step-response}Step response of the filter in Eq. \ref{eq:discrete inversion}
shows how the different components of hybrid inversion combine to
reconstruct the wavelength shift signal. The source wavelength (black)
is a discrete Heaviside function with step at pixel 100. The differential
stereoscopy step (in the absence of noise) yields a high-pass filtered
version of the Heaviside function (blue), while the correlation step
yields a low-pass filtered version (cyan). Because the filters are
constructed mathematically to match one another, the sum of the two
filters (magenta) reproduces the step. The low-pass curve suffers
from edge effects at the far sides of the image, but in a typical
detector with more than $10^{3}$ pixel width, edge effects are minimal.}
\end{figure}

Figure \ref{fig:step-response} shows the step response of the filter
described by Eq. \ref{eq:discrete inversion}, to illustrate how the
different parts of the inversion work together. The step is located
at position 100. The high-pass response to the differential stereoscopy
signal (dark blue) is due to the interaction between the differentiation
inherent in the detection method, and the {}``imperfect'' (high-pass)
integration imposed by the $\epsilon$$\Lambda'_{c1}$ decay term
in the integrand of Eq. \ref{eq:hybrid_inversion}. The exponential
decay is matched by the low-pass response to the correlation stereoscopy
signal (light cyan). The sum of the two signals is the value of the
the RHS of Eq. \ref{eq:discrete inversion}, and reproduces the original
signal. The slight overshoot of the reconstructed (magenta) curve
is a discretization effect due to the phase of the Heaviside function
relative to the pixel grid: a Heaviside function with the opposite
phase relative to the pixel grid would yield an equal-magnitude undershoot.
As with any discrete sampling filter, Eq. \ref{eq:discrete inversion}
operates best on spatial frequencies less than 2/3 of the pixel Nyquist
frequency, because the reconstructed amplitude of those frequency
components is independent of their phase relative to the pixel grid.
The Heaviside function contains frequencies both below and above that
limit, and the high frequencies' reconstructed amplitude depends on
phase.

In practice, errors in both the measurement of intensity and the inversion
affect the analysis, so that the measured $\Lambda'_{c1}$ is not
exactly the same as $\Lambda'$ even if the image frequency spectrum
is appropriately limited. In this error analysis we consider six principal
sources of error: photon counting noise, $\left\langle \Delta I\right\rangle_{RMS} $,
that represents an additive, uncorrelated noise source at each pixel
in each of the differential and correlation stereoscopy steps; misalignment
noise $\left\langle \Delta X_{L}\right\rangle_{RMS} $ in the cross-correlation
step; the effect of intensity gradients (both $dC/dx$ and $dL/dx$)
on the line shift signal; a nonzero slope $dF/d\lambda'$ of the instrument
filter transmission function $F(\lambda')$ (the slope is assumed
to be zero in Eq. \ref{eq:hybrid_inversion}); and an excessively
high $d\Lambda'/dx$ signal in the solar image. The first two noise
sources arise (directly or indirectly) from imprecision in the measurement
of $I$ in each pixel, and the last four sources arise from the non-ideal
nature of the instrument and/or the solar field being imaged. In the
following subsections we treat each error source independently.

\subsection{Photon noise in differential stereoscopy\label{sub:Photon-counting-noise}}

Photon counting noise is due to uncertainty in the measured intensity
$I_{\pm1}$ in each focal plane, through the statistics of photon
counting. Each measurement does not determine intensity directly,
but rather fluence of energy (or, equivalently, photon count) on each
pixel during an exposure time. Intensity can be derived by knowing
the size of the pixel and the length of the exposure. Each pixel measures
a fluence $\Phi$ with an uncertainty $\Delta\Phi_{ph.}$ given by:\begin{equation}
\Delta\Phi_{ph.}=\frac{hc}{\Lambda}\Delta n_{ph.}=\sqrt{\frac{hc}{\Lambda}\Phi}\label{eq:photon-noise}\end{equation}
 so that the RMS value $\left\langle \Delta I/I\right\rangle_{RMS} =\left\langle \Delta\Phi/\Phi\right\rangle_{RMS} =\sqrt{hc/\Lambda\Phi}$.
This effect contributes to $\Lambda'_{c1}$ primarily via a photon
noise term that must be propagated through Equation \ref{eq:discrete inversion}:
the much higher photon counts in many-pixel image {}``patches''
greatly reduce photon counting noise in the cross-correlation portion
of the inversion (\S \ref{sub:Misalignment-noise}). The RMS photon
noise contribution $\left\langle \Delta_{ph}\Lambda'_{c1}\right\rangle_{RMS} $
is just the incoherent sum of the photon noise terms in each term
on the RHS of Eq. \ref{eq:discrete inversion}. Each of those terms,
in turn, is (to first order in $\Delta_{ph}I/I$ and neglecting $dI/dx$
for the calculation of noise in each pixel):\begin{equation}
\left\langle \Delta_{ph}J_{i}\right\rangle_{RMS} =\sqrt{\frac{2I^{2}\left(hc\Phi/\Lambda\right)}{4I^{2}}}=\sqrt{\frac{hc}{2\Lambda\Phi}}=\sqrt{1/2n_{ph}}\label{eq:delta_ph J}\end{equation}
where $n_{ph}$ is the per-exposure photon count in the corresponding
pixel in each of the two orders. Eq. \ref{eq:delta_ph J} simply reproduces
the familiar behavior of incoherent counting statistics for intrinsic
(normalized) quantities. 

The treatment may, at first glance, appear overly simple as the individual
pixels in the images are interpolated and hence do not represent individual
samples of the $\Delta_{ph}\Phi$ random variable but rather scaled
incoherent sums of adjacent samples. This is acceptable to ignore
because, in the absence of strong gradients in $dX_{L,\pm1}/dx$ (i.e.
gradients comparable to unity per pixel), the samples' noise profile
is preserved by the inversion. Adjacent pixels are co-added in the
discrete inversion. Under linear interpolation, adjacent interpolated
fractions of the signal from each detector pixel are approximately
conserved by the interpolation and added coherently in the cross-pixel
sum in Eq. \ref{eq:discrete inversion}. Because addition is commutative,
the noise effect of the interpolated pixels on the final sum is the
same as if no interpolation had taken place, up to a negligible factor
of $(1-\epsilon$). Put another way, although considering interpolation
would further reduce the noise in $J_{i}$ for individual pixels,
co-adding $J_{i}$ for adjacent pixels in the inversion eq. \ref{eq:discrete inversion}
would then include coherent cross-pixel noise terms that would exactly
cancel the reduction -- so we may safely choose, instead, to ignore
the effects of the $X_{L,\pm1}$ interpolation and use incoherent
sums throughout.

Carrying Eq. \ref{eq:delta_ph J} forward through Eq. \ref{eq:discrete inversion}
yields:\begin{equation}
\left\langle \Delta_{ph}\Lambda'_{c1}\right\rangle_{RMS} =\frac{1}{2\alpha}\left\langle \frac{C}{L}\right\rangle_{RMS} \sqrt{\left(\frac{1}{2n_{ph}}\right)\left(\sum_{n=0}^{i}\left(1-\epsilon\right)^{2(i-n)}+\sum_{n=i}^{w-1}\left(1-\epsilon\right)^{2(n-i)}\right)}\label{eq:ph-noise-1}\end{equation}
Neglecting any field edge effects (in other words, taking $0\ll i\ll w$)
permits taking the infinite sums directly, yielding (to first order
in epsilon):\begin{equation}
\left\langle \Delta_{ph}\Lambda'_{c1}\right\rangle_{RMS} =\frac{1}{2\alpha}\left\langle \frac{C}{L}\right\rangle_{RMS} \sqrt{\frac{1}{2n_{ph}\epsilon}}\label{eq:ph-noise}\end{equation}
Thus the statistics of the differential stereoscopy step are worsened
by the width of the correction term. This is reflective of the fact
that random walks diverge, and is the reason for using a hybrid inversion
rather than a direct integral inversion: using direct integration
in Equation \ref{eq:discrete inversion}, rather than high-pass integration
with correction, is equivalent to setting $\epsilon$ to zero, which
would yield an arbitrarily large amount of photon-counting noise in
the individual measurements. In actual use, it is necessary to balance
$\epsilon$ and the cross correlation patch size: larger patch sizes
improve correlation behavior but require smaller $\epsilon$ and thus
worsen the photon noise due to integration of the differential signal.

It is important to note that the photon noise from the discrete inversion
is not independently sampled at each pixel: it is the sum of many
noise terms along the $x$ direction, so that the \emph{difference}
measurement between adjacent pixels has a much smaller noise level,
yielding higher precision for small features. In particular, shifting
by 1 pixel re-samples only one $\epsilon^{th}$ of the population
of random samples in the calculated $\Lambda'_{c1}$, reducing the
photon noise by a factor of $\sqrt{\epsilon}$ so that small features
are only subject to: \begin{equation}
\left\langle \Delta_{ph}\left(\Lambda'_{c1,i}-\Lambda'_{c1,j-1}\right)\right\rangle_{RMS} =\frac{1}{2\alpha}\left\langle \frac{C}{L}\right\rangle_{RMS} \sqrt{\frac{1}{2n_{ph}}},\label{eq:differential-ph-noise}\end{equation}
 which is to be expected considering that the difference signal is
the original signal being measured by the differential stereoscopy
- the difference in measured wavelength between two adjacent pixels
is simply a linearly scaled form of $J_{i}$. Differential stereoscopy
is \emph{more sensitive} to small features than to large ones, at
least from the standpoint of photon statistics. 

Taking {}``typical'' values of $\epsilon=0.09$ (FWHM of 15 pixels),
$C/L=10$, and $n_{ph}=5\times10^{4}$ yields an absolute partial
sensitivity $\alpha\left\langle \Delta_{ph}\Lambda'_{c1}\right\rangle_{RMS} =0.052$
pixel, and a gradient sensitivity $\alpha\left\langle \Delta_{ph}\left(\Lambda'_{c1,j}-\Lambda'_{c1,j-1}\right)\right\rangle_{RMS} =0.016$
(the gradient sensitivity is unitless: pixels per pixel), due to photon
counting alone.

\subsection{Photon noise in correlation stereoscopy\label{sub:Misalignment-noise}}

The second source of error in Equation \ref{eq:discrete inversion}
is misalignment in the cross correlation step that is used to derive
$X_{L,\pm1}$ from the original images. In the 2004 derivation of
Eq. \ref{eq:hybrid_inversion}, it was necessary to assume that the
displacement of the images was small, leading to $dI/dx$ terms that
enter in first order because the +1 and -1 order intensities are measured
at slightly different spatial locations. Here, we consider the effect
of small misalignments due to errors in the correlation co-alignment
process. In the test measurements made by DeForest et al., the cross-correlation
step used a full two-dimensional patch of image around each sampled
point, rather than (as in the differential stereoscopic portion of
the inversion) merely the particular horizontal line surrounding each
point. This is reasonable in part because the image is, in fact, a
two-dimensional field rather than a collection of independent rasters;
and in part because the technique was found empirically to reduce
correlation {}``misses'' and improve the image signal. Here, we
consider the potential sources of error injected by cross-correlation
alignment of patches of image around each point.

Because correlation co-alignment is a nonlinear process and depends
on the field in the image, it is necessary to make some assumptions
about the field being viewed. We here confine our analysis to magnetic
measurements of the quiet Sun, which consists principally of nearly
uniformly bright granules separated by dark, well defined lanes. Some
other types of field (for example, the UV chromospheric network) are
analogous and may be considered with the same or similar analysis,
but other types of field (such as sunspot penumbrae) are likely too
complex for simple analysis and require numerical modeling.

\begin{figure}
\includegraphics[width=3in]{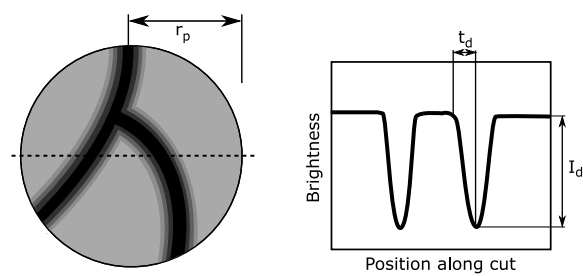}

\caption{\label{fig:correlation}Sample correlation field used for our correlation
photon analysis. Correlation between two images is used to align two
patches of pixels with radius $r_{p}$. The image field consists of
dark lanes against a background of approximately uniform brightness
(left). The lanes consist of side-walls with thickness $t_{d}$ and
depth $I_{d}$ (right). A fraction $f_{d}$ of the image is considered
to be lane wall.}
\end{figure}

In this analysis we consider cross-correlation of an idealized small
patch of quiet Sun with $\pi r_{p}^{2}$ pixels, of nearly uniform
intrinsic brightness everywhere except in a fraction $f_{d}$ of the
pixels of which are occupied by narrow, dark features with steep sides
that are, on average, $t_{d}$ pixels across, and an absolute intensity
deficit of $I_{d}$ (Figure . For typical quiet-Sun visible light
observations, $f_{d}$is about 0.1 and $I_{d}$is about $0.5\, C$.
Assuming that the correlation algorithm does not 'miss' (it achieves
a match between corresponding features in the opposite order images),
we assume that errors in the resulting overall position of the patches
is due to randomization errors from the displacements (real and apparent)
of the dark features between the two images. In the worst non-pathological
case, every intensity feature in the image patch is displaced by a
different random amount, representing a strong high-spatial-frequency
field. In that case, each feature's apparent location is the incoherent
sum of two random variable samples: one due to photon counting statistics
in the sloping sides of the feature, and one due to the random displacements
of the features themselves. The former depends on the steepness of
feature sides, the overall brightness of the image, and the number
of pixels in each feature wall. The latter depends on the dispersion
of the instrument and the variability of spectral offset across the
image

The uncertainty in location of each feature due to photon counting
statistics is related to the slope at the sides of the feature: correlation
fitting converts intensity in the sloping sides of a feature to position,
so we can estimate the position/wavelength noise due to error in the
intensity, by simply scaling the intensity noise to account for this
conversion. The cross-correlation requires comparison between two
features, so one can write:\begin{equation}
\left\langle \Delta X_{ph,1}\right\rangle_{RMS} =\frac{t_{d}}{I_{d}}\left\langle \Delta I_{ds}\right\rangle_{RMS} =\frac{t_{d}}{I_{d}}I_{0}\sqrt{\frac{2}{t_{d}n_{ph}}\left(\frac{2I_{0}}{2I_{0}-I_{d}}\right)}\label{eq:single-feature}\end{equation}
where $\Delta I_{ds}$ is the uncertainty in brightness in the side-of-feature
pixels, and the right-hand expression accounts for the typical number
of pixels in the sides of a particular feature. The $n_{ph}$ comes
from the fraction $\left\langle \Delta I\right\rangle_{RMS} /I_{0}$and
is the number of photons counted during the exposure. The unitless
$2I_{0}/(2I_{0}-I_{d})$ arises because the average brightness in
a feature sidewall is halfway between the background intensity and
the dark feature intensity, The extra factor of 2 under the radical
comes from the fact that the noise in offset is the incoherent sum
of the location noise in each of the two features being compared. 

The ensemble average error position $\left\langle \Delta X_{ph}\right\rangle_{RMS} $
is then just the incoherent average of the offsets of each feature
wall, or \begin{equation}
\left\langle \Delta X_{ph}\right\rangle_{RMS} =\frac{2I_{0}}{I_{d}r_{p}}\sqrt{\frac{I_{0}t_{d}}{\pi f_{d}n_{ph}\left(2I_{0}-I_{d}\right)}}\label{eq:correlation-photon-noise}\end{equation}
 Taking the patch to be about 15 pixels across, and assuming a scene
of solar granulation with $f_{d}=0.05$, $t_{d}=2$, $I_{d}=0.5I_{0}$,
and $n_{ph}=5\times10^{4}$ yields a noise estimate of $\left\langle \Delta X_{ph}\right\rangle_{RMS} =2.4\times10^{-3}$
pixels. Photon noise in the correlation step is hence negligible compared
to photon noise in the differential step. This is not surprising considering
the large number of pixels that are averaged over by the correlation
step. Put another way, the noise spectrum is dominated by high spatial
frequencies, which are {}``smoothed out'' by the correlation process.

\subsection{\label{sub:Image-noise-in}Image noise in correlation stereoscopy}

There is an additional source of error in the correlation step: because
the wavelength signal $\Lambda'(x)$ includes high spatial frequencies
(the signal that is being measured by the differential stereoscopy
step), different parts of the image must be shifted in different directions
to find the average offset over the entire patch being correlated.
These high spatial frequencies introduce an error into the measurement
of the shifts $X_{L,\pm1}(x)$, because the correlation step weights
the importance of the different offsets according to the average sidewall
slope in each portion of the patch being correlated, rather than uniformly
throughout the patch of image. This error term is independent of the
photon noise, and is introduced even for successful cross-correlations.
It depends on the inequality of the weighting of the features, as
well as the amount of displacement:\begin{equation}
\left\langle \Delta X_{wgt}\right\rangle_{RMS} =\alpha\frac{\left\langle \overline{\delta\Lambda'_{patch}\delta\left|dI/dx\right|}\right\rangle_{RMS} }{\left\langle dI/dx\right\rangle_{RMS} }\label{eq:weighting-delta}\end{equation}
where $\delta$ is used to indicate actual differences from the patch-wide
mean (different from $\Delta,$which is being used to indicate statistical
error in a quantity), $\Lambda'_{patch}$ is the signed difference
between $\Lambda'$ at each location in the patch being correlated
and the mean for the whole patch, $dI/dx$ is the slope of the intensity
function at each location in the image, and $\delta\left|dI/dx\right|$
is the difference between the current value of the slope and its patch-wide
average. As above, angle-brackets symbolize RMS averaging over a complete
data set; here, an over bar is used to represent signed averaging
over the patch. The inclusion of both $\delta$-quantities (on the
RHS) and $\Delta$-quantities (on the LHS) is because Eq. \ref{eq:weighting-delta}
describes errors in the measurement of $X$ that are due to the interpretation
of the images themselves even in the absence of injected noise, rather
than to an injected noise source in the direct measurement of brightness
in each image plane.

The numerator in the RHS of Eq. \ref{eq:weighting-delta} depends
only on the scene being viewed. The signed average simply applies
an $n_{samp}^{-1/2}$ scaling of the overall error term, according
to the number of independent samples $n_{samp}$ of image data in
the patch. The quantity being averaged is the product of two random
variables with zero mean and hence can be expanded simply in terms
of their covariance $C(x,y)$:\begin{equation}
\left\langle \Delta X_{wgt}\right\rangle_{RMS} =\frac{\alpha}{\left\langle dI/dx\right\rangle_{RMS} \sqrt{n_{f}}}\left(C\left(\delta\Lambda'_{patch},\,\delta\left|dI/dx\right|\right)+\left\langle \delta\Lambda'_{patch}\right\rangle_{RMS} \left\langle \delta\left|dI/dx\right|\right\rangle_{RMS} \right)\label{eq:weighting-covariance}\end{equation}
where $n_{f}$ is the number of {}``features'' -- separately resolved
regions that each represents an independent sample of image data --
in the patches being correlated. 

Assuming that the wavelength and brightness signals are uncorrelated
eliminates the covariance term:\begin{equation}
\left\langle \Delta X_{wgt,uncorr}\right\rangle_{RMS} =\frac{\alpha\left\langle \delta\Lambda'_{patch}\right\rangle_{RMS} \left\langle \delta\left|dI/dx\right|\right\rangle_{RMS} }{\left\langle dI/dx\right\rangle_{RMS} \sqrt{n_{f}}}\label{eq:no-covariance}\end{equation}
so that, if wavelength and intensity of emissions be uncorrelated,
the location uncertainty of a given cross-correlated patch is proportional
to the RMS variation in central wavelength and the RMS variation in
\emph{slope} of the image intensity. For example, taking the patch
to be about 15 pixels across, $n_{f}=n_{p}/4$ (one unique feature
every four pixels), $\alpha\left\langle \delta\Lambda'_{patch}\right\rangle_{RMS} $=
0.1 pixel, $\left\langle dI/dx\right\rangle_{RMS} $=0.2, and $\left\langle \delta\left|dI/dx\right|\right\rangle_{RMS} $=0.1
yields $\left\langle \Delta X_{wgt,uncorr}\right\rangle_{RMS}=0.006$~pixel. 

In practice, the image intensity and wavelength variation are not
completely uncorrelated. For example, small magnetic features tend
to be found in dark intergranular lanes, rather than randomly across
the quiet Sun, so that $C\left(\delta\Lambda'_{patch},\,\delta I\right)\neq0$.
It is not clear, however, how or whether $dI/dx$ and $\Lambda'$
are correlated. In the worst case, they are proportional and the covariance
term is equal to the product term in Eq.\ref{eq:weighting-covariance},
doubling the effective noise level from this source to (in the example)
0.012 pixel. In the best case, $\delta\Lambda'$ would be near zero
everywhere except where $dI/dx$ is small. In that case, the covariance
term would nearly exactly cancel the product term in Eq. \ref{eq:weighting-covariance}
and eliminate this noise source.

In the context of a photospheric magnetograph, some further information
about the covariance term may be used. In the quiet Sun for example,
the strongest magnetic fields are observed to exist in the intergranular
lanes, near local minima in continuum intensity. This suggests that
the covariance between $dI/dx$ and $\delta\Lambda'_{patch}$ is in
fact negative in real scenes of solar granulation, so that Eq. \ref{eq:no-covariance}
likely overestimates the strength of this noise source in quiet sun
magnetic measurements. 

In scenes that contain more contrast, particularly magnetic-correlated
contrast, image correlation problems can dominate the Zeeman shift
signal. in particular, near sunspots very strong variations in intensity
are strongly correlated to very strong variations in magnetic field,
reducing the effectiveness of this technique for inverting stereoscopic
data of sunspot fields. This was observed by DeForest et al. (2004),
who found that correlation introduced strong artifacts into a not-well-resolved
sunspot penumbra.

\subsection{\label{sub:Image-brightness-gradients}Image brightness gradients}

Equation \ref{eq:hybrid_inversion} includes several important engineered
cancellations. Failure of those cancellations causes leakage of intensity
or other information into the derived wavelength signal. The most
important, and most likely to fail, cancellation is the elimination
of an intensity spatial-derivative signal from the $J$ introduced
in Eq. \ref{eq:J-definition}. In the narrow-line approximation described
by DeForest et al. (2004), the single-image brightness is given by
(their equation 7):\begin{equation}
I_{n}\approx\frac{\left[E_{c}\otimes F\right](x/\alpha_{n})}{\alpha_{n}}-L(x)+\alpha_{n}L(x)\frac{d\Lambda'}{dx}+\alpha_{n}\Lambda'(x)\frac{dL}{dx}-2\alpha_{n}^{2}\Lambda'(x)\frac{dL}{dx}\frac{d\Lambda'}{dx}\label{eq:single-intensity}\end{equation}
(where, again, $E_{c}$is the continuum brightness per unit wavelength,
$F$ is the prefilter function, $L(x)$ is the total line depth intensity,
and $\alpha_{n}$ is the dispersion of the $n^{th}$ spectral order).
The differential inversion arises from noticing that the difference
between two opposite orders contains a useful $d\Lambda'/dx$ term:
\begin{equation}
I_{n}(x)-I_{-n}(x)=2\alpha_{n}L(x)\frac{d\Lambda'}{dx}+2\alpha_{n}\Lambda'(x)\frac{dL}{dx}\label{eq:intensity-equation}\end{equation}
in which the $d\Lambda'/dx$ term is desired and the $dL/dx$ term
is undesired. In hybrid stereoscopy the images are resampled into
$X_{L}$ coordinates with the substitution: \begin{equation}
X_{L,n}=x+\alpha_{n}\Lambda'_{L}(x)\label{eq:X-def}\end{equation}
 that eliminates the undesired $dL/dx$ term in Eq. \ref{eq:intensity-equation}
to first order. This process inevitably introduces some other undesired
terms. In particular, substituting Eq. \ref{eq:X-def} into Eq. \ref{eq:single-intensity}
yields (writing $C_{L}(x)$ for the convolution term):

\begin{eqnarray}
I_{n}(X_{L,n}) & \approx & \left(C_{L}(x)+\alpha_{n}\Lambda'_{L}\frac{dC}{dx}\right)-\left(L(x)+\alpha_{n}\Lambda'_{L}\frac{dL}{dx}\right)\nonumber \\
 &  & +\left(\alpha_{n}L\frac{d\Lambda'}{dx}+\alpha_{n}^{2}\Lambda'_{L}\frac{dL}{dx}\frac{d\Lambda'}{dx}+\alpha_{n}^{2}\Lambda'\Lambda'_{L}\frac{d^{2}L}{dx^{2}}\right)\nonumber \\
 &  & +\left(\alpha_{n}\Lambda'\frac{dL}{dx}+\alpha_{n}^{2}\Lambda'\Lambda'_{L}\frac{d^{2}L}{dx^{2}}+\alpha_{n}^{2}\Lambda'_{L}\frac{d\Lambda'}{dx}\frac{dL}{dx}\right)\nonumber \\
 &  & -\left(2\alpha_{n}^{2}\Lambda'\frac{dL}{dx}\frac{d\Lambda'}{dx}+2\alpha_{n}^{3}\Lambda'_{L}\left(\left(\frac{d\Lambda'}{dx}\right)^{2}\frac{dL}{dx}+\Lambda'\frac{d^{2}L}{dx^{2}}\frac{d\Lambda'}{dx}+\Lambda'\frac{dL}{dx}\frac{d^{^{2}}\Lambda'}{dx^{2}}\right)\right)\label{eq:eviltude}\end{eqnarray}
where $C(x)$ is the convolution term from Eq. \ref{eq:single-intensity}.
As before, subtracting opposite orders removes all terms that contain
an even power of $\alpha$, and doubles the remaining terms. Grouping
all terms and discarding terms beyond first order in $\Lambda'$ yields:\begin{equation}
I_{n}(X_{L,n})-I_{-n}(X_{L,-n})\approx2\alpha_{n}\left(L\frac{d\Lambda'}{dx}+\Lambda'_{L}\frac{dC_{L}}{dx}+\Lambda'_{H}\frac{dL}{dx}-2\alpha_{n}^{2}\left(\frac{d\Lambda'}{dx}\right)^{2}\Lambda'_{L}\frac{dL}{dx}\right)\label{eq:not-so-evil}\end{equation}
where the first term is desired and the other terms are noise. The
$dC_{L}/dx$ term is normally small both because the filter function
is symmetric in wavelength and because the convolution integral smooths
across several pixels. The $\Lambda'_{H}$ term represents leakage
of the $dL/dx$ signal into the data. (Recall, $\Lambda'_{H}$ is
just the high spatial frequency component of the wavelength shift,
given by $\Lambda'-\Lambda'_{L}$). The low-frequency image contribution
to the noise has been reduced by the square of the (normally small)
factor $\alpha_{n}d\Lambda'/dx$.

The LHS of Equation \ref{eq:not-so-evil} is merely the numerator
of the $J$ definition in Equation \ref{eq:J-definition}. The denominator
is the sum of the two intensities, which retains the even powers of
$\alpha_{n}$ while canceling the odd powers. Again, many cancellations
occur, leaving:\begin{equation}
I_{n}(X_{L,n})+I_{-n}(X_{L,n})\approx2\left(C_{L}(x)-L+2\alpha_{n}^{2}\Lambda'\Lambda'_{L}\frac{d^{2}L}{dx^{2}}\right)\label{eq:non-evil}\end{equation}
so that the summed brightness of the resampled images has a systematic
noise term that is proportional to the second derivative of the line
depth. Combining the terms yields a more complete expression for $J$:\begin{eqnarray}
J & \approx & \alpha_{n}\frac{L\frac{d\Lambda'}{dx}+\Lambda'_{L}\frac{dC_{L}}{dx}+\left(\Lambda'_{H}-\left(\alpha_{n}\frac{d\Lambda'}{dx}\right)^{2}\Lambda'_{L}\right)\frac{dL}{dx}}{\left(C_{L}(x)-L\right)+2\alpha_{n}^{2}\Lambda'\Lambda'_{L}\frac{d^{2}L}{dx^{2}}}\nonumber \\
 &  & \approx\alpha_{n}\frac{L}{C_{L}}\left(\frac{d\Lambda'}{dx}+\Lambda'_{L}\frac{dC_{L}}{L\, dx}+\left(\Lambda'_{H}-\left(\alpha_{n}\frac{d\Lambda'}{dx}\right)^{2}\Lambda'_{L}\right)\frac{dL}{L\, dx}-\frac{2\alpha_{n}^{2}\Lambda'\Lambda'_{L}}{C_{L}}\frac{d^{2}L}{dx^{2}}\frac{d\Lambda'}{dx}\right)\label{eq:evil-J}\end{eqnarray}
where first order approximations to the fraction have been taken,
and squares of perturbations have been ignored. The first term of
Eq. \ref{eq:evil-J} reproduces the desired signal (Eq. \ref{eq:dlambda/dx})
and the additional terms represent noise sources due to the image
itself. Writing them as an error term $\Delta J_{grad}$,\begin{equation}
\Delta J_{grad}\approx\frac{\alpha_{n}\Lambda'_{L}}{C_{L}}\left(\frac{dC_{L}}{dx}+\left(\frac{\Lambda'_{H}}{\Lambda'_{L}}-\left(\alpha_{n}\frac{d\Lambda'}{dx}\right)^{2}\right)\frac{dL}{dx}-2\alpha_{n}\Lambda'\frac{d^{2}L}{dx^{2}}J_{ideal}\right)\label{eq:deltaj}\end{equation}
 where the first terms represent leakage of intensity information
into the wavelength signal (both from continuum variation and from
line variation), and the final term represents leakage of rapid brightness
variations into the J gain and is negligible in nearly all cases. 

The continuum intensity leakage term is dependent on large-scale variations
in continuum brightness, because $C_{L}$is smoothed (by convolution
with the filter function) over a distance of $\alpha_{n}\delta\lambda$,
where $\delta\lambda$ is the width of the entrance filter. In typical
applications, $\alpha_{n}\delta\lambda$ might be 5-15 pixels. A scene
of solar granulation with $C/L$\textasciitilde{}10, $\alpha\Lambda'_{L}$\textasciitilde{}0.1
pixel and $\alpha_{n}\delta\lambda$\textasciitilde{}10 pixels, with
intergranular lanes 3 pixels wide and continuum dips of 30\% in the
lanes, will have a continuum intensity leakage error of \textasciitilde{}$5\times10^{-4}$
in J/I in the regions on either side of the lane, equivalent to the
noise from $10^{6}$ photons. The error accumulates coherently over
half of the dispersed filter width, or about 5 pixels, yielding a
peak injected noise signal in $\alpha\Lambda'$ of \textasciitilde{}0.025
pixel in the lane center from just the continuum term.

The continuum intensity leakage is much stronger in regions with strong
large scale variations in intensity, coupled with strong line shifts
(such as found in sunspots). In a sunspot scene with $C/L$\textasciitilde{}10,
$\alpha\Lambda'_{L}\sim4$ pixels, and a continuum dip of 75\% within
the sunspot, pixels near the sunspot edge will have a continuum leakage
term of 0.2 in J/I, much larger than any photon noise in the system.
The error accumulates coherently for the full dispersed width of the
entrance filter, yielding integrated errors in $\alpha\Lambda'_{L}$
as high as 20 pixels near the edge of the sunspot. This large error
injection is the primary reason why DeForest et al. (2004) found sunspot
measurements to be challenging with this inversion technique.

The line intensity leakage term is largest in regions of very rapid
variation of the central wavelength, but is negligible compared to
the continuum leakage in most scenes. Considering a {}``challenging''
scene of solar plage with $\alpha\Lambda'_{L}$\textasciitilde{} 1
pixel, $\alpha\Lambda'_{H}\sim0.3\, pixel$ amplitude, $\alpha d\Lambda'/dx$\textasciitilde{}
0.1, $C/L\sim10$, and $dL/Ldx\sim0.5/pixel$ yields intensity leakage
of about $7\times10^{-3}$ into J/I from the central term. This is
equivalent to the photon noise from $3\times10^{4}$photons, and if
sustained across a full correlation patch of width 15 pixels could
inject an error signal of up to \textasciitilde{}0.1 pixel into the
final inversion; this represents about 20\% of the background field.
With a more typical quiet-Sun case of $\alpha\Lambda'_{L}<0.1$pixel
(30 G), $\alpha\Lambda'_{H}\sim0.1$pixel, and $\alpha d\Lambda'/dL\sim0.02$,
the intensity leakage noise is under $10^{-3}$ in J/I and thus leads
to errors of under 0.015 pixel after accumulating over a 15-pixel
FWHM patch.

The rapid brightness variation leakage depends on rapid changes in
the line depth, in regions that are magnetized. It is negligible in
nearly all cases other than sunspot penumbrae. In the quiet Sun, the
strongest value of the term occurs where $g$-band bright points are
present in intergranular lanes. In a typical g-band bright point,
with $\alpha\Lambda'\sim1\, pix$, $\alpha\Lambda'_{L}\sim0.1\, pix$and
$d^{2}L/dx^{2}\sim0.005I\, pix^{-2}$, $J$ can be expected to have
a relative error of only $1\times10^{-3}$ from this term. In sunspot
penumbrae, where both $\alpha\Lambda'_{L}$ and $\alpha\Lambda'$
may be larger than order unity, and where $d^{2}L/dx^{2}$ may be
a factor of 4-10 higher than in intergranular lanes due to the fine
nature of penumbral roll structure, $J$ can be expected to have a
relative error of order close to unity from this term, severely degrading
results in sunspot penumbrae even if the continuum leakage signal
(above) were not a problem.

In all cases, the intensity gradient leakage terms are proportional
to $\Lambda$' and therefore represent a gain error rather than an
uncorrelated additive noise source that affects detection of lone
magnetic features.

\subsection{\label{sub:Filter-passband-gradients}Filter passband gradients}

Filter passband gradients enter because the overall measured intensity
is affected by the throughput function of the prefilter. In conventional
filtergraph instruments, the slope of the side-lobes of the main filter
passband is used to convert wavelength offset into an intensity signal;
in stereographic instruments, that signal is undesired and represents
a source of noise. Even in a properly tuned stereoscopic instrument,
large excursions of $\Lambda'$ carry the spectral line out of the
flat central region of the prefilter's passband and into the sloping
wings. This modifies Eq. \ref{eq:single-intensity}, to include the
effect of the filter on the measured L(x):\begin{eqnarray}
I_{n} & \approx & \left\{ \begin{array}{c}
\frac{\left[E_{c}\otimes F\right](x/\alpha_{n})}{\alpha_{n}}-L(x)F(\Lambda')+\alpha_{n}L(x)F(\Lambda')\frac{d\Lambda'}{dx}\\
+\alpha_{n}\Lambda'(x)\left(F(\Lambda')\frac{dL}{dx}+L(x)\frac{dF}{d\Lambda'}\frac{d\Lambda'}{dx}\right)\\
-2\alpha_{n}^{2}\Lambda'(x)\left(F(\Lambda')\frac{dL}{dx}\frac{d\Lambda'}{dx}+L(x)\frac{dF}{d\Lambda'}\left(\frac{d\Lambda'}{dx}\right)^{2}\right)\end{array}\right\} \label{eq:asymmetric-filter}\end{eqnarray}
where $F(0)$ is taken to be unity. Of the two new terms, only one
is asymmetric, so the difference equation becomes:\begin{equation}
I_{n}-I_{-n}=F(\Lambda'(x))\left(2\alpha_{n}L(x)\left(1+\frac{\Lambda'}{F(\Lambda'(x))}\frac{dF}{d\Lambda'}\right)\frac{d\Lambda'}{dx}+2\alpha_{n}\Lambda'(x)\frac{dL}{dx}\right)\label{eq:asymmetric-diff}\end{equation}
where both the overall factor of F and the additional unitless derivative
$(\Lambda'/F)(dF/d\Lambda')$ represent perturbations on the gain
of the measured $d\Lambda'/dx$, and the final term is the intensity
leakage that is removed by the hybridization step in the overall inversion.
In a well-tuned instrument, the maximum value of $F$ occurs near
$\Lambda'=0$, so we may expand F and F' around that value: \begin{equation}
F\approx1-\frac{1}{2}\left(\frac{\Lambda'-\Lambda'_{filt}}{W}\right)^{2}\label{eq:filter}\end{equation}
where $\Lambda'_{filt}$ is the difference between the center wavelength
of the filter and the rest wavelength of the spectral line, and $W$
is the half-width at half maximum of the filter passband. Then the
derivative $dF/d\Lambda'$ is given by $(\Lambda'_{filt}-\Lambda')/W$.
As an example, if $\Lambda'_{filt}$ is about -W/10 and $\Lambda'$is
about W/4, this yields a gain error of 7\% in the $d\Lambda'/dx$
signal.

\subsection{\label{sub:Rapid-variation-of}Rapid variation of central wavelength}

Rapid variation of the central wavelength of the observed spectral
line affects the measured signal from differential stereoscopy. The
differential inversion (Eq. \ref{eq:discrete inversion}) is derived
using a first-order expansion of the brightness in terms of the slope
$d\Lambda'/dx$ of the central wavelength versus position. If the
condition $d\Lambda'/dx\ll\alpha_{n}^{-1}$ does not hold (the slope
is significant compared to the reciprocal of the instrument's dispersion)
then the differential inversion equation is not valid and the derived
$\Lambda'_{c1}$ will differ from the true $\Lambda'$ even in the
absence of any other noise source. This is due to the effect being
exploited to extract the $d\Lambda'/dx$ signal from the intensities:
small shifts in the $\lambda$ direction project into the dispersed
spatial direction, so that the line of integration in the $(x,\lambda)$
plane that contributes to a single pixel may have more or less contribution
from the line core depending on the relative angle between $\alpha_{n}$
and $d\Lambda'/dx$. The resultant variation in intensity is extremely
nonlinear when $\alpha_{n}$ and $d\Lambda'/dx$ are nearly parallel.
Further, the $d\Lambda'/dx$ intensity variations are due to small
shifts in image position as$\Lambda'$ varies. If $\Lambda'$ grows
larger than 1 pix/$\alpha_{n}$, then the image shift becomes noticeable
and distorts the resultant line-shift images. 

\begin{figure}
\includegraphics[width=4.0in]{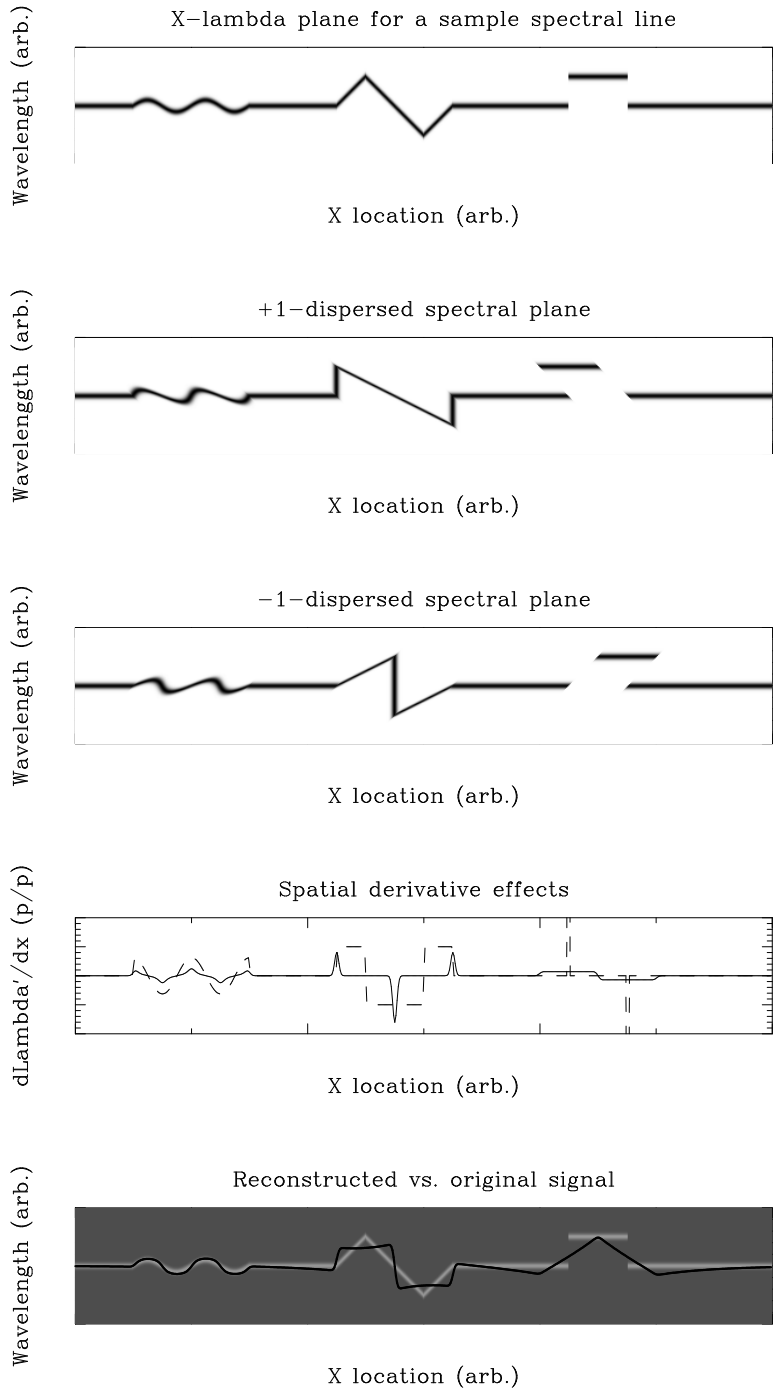}

\caption{\label{fig:multi}Five plots showing arbitrary variations of spectral
line central wavelength show the manner in which the approximation
of Equation \ref{eq:discrete inversion} breaks down at high line
slopes. In order: the X-Lambda plane; the X-lambda plane, dispersed
in two opposite directions (Pixel brightnesses are formed by vertical
integration in these two plots); the original and derived spatial
derivative of line center wavelength; and an overlay of the original
and reconstructed signals. The sinusoidal segment at left is well
matched. The pathological sawtooth and square wave segments are distorted
by the coupling between wavelength and spatial position.}
\end{figure}

The various effects of large excursions in $\Lambda'$ on spatial
distributions are illustrated in Figure \ref{fig:multi}, which illustrates
the geometry of the $(x,\lambda)$ plane in a stereoscopic measurement
of line center for a spectral line with structured Doppler signal.
Three forms of perturbation on the line are shown: a sinusoidal modulation
of the line center wavelength; a triangle wave modulation; and a dual
step function. The sinusoidal modulation fits the criteria outlined
by DeForest et al. (2004) for differential stereoscopic inversion:
$\alpha_{n}d\Lambda'/dx\gg1$ and $\alpha_{n}\Lambda'$ is not large
compared to the spatial features of interest. The others fail in different
ways. The triangle wave fails because $\alpha_{n}d\Lambda'/dx\sim1$,
and the offset fails because $\alpha_{n}\Lambda'$is large compared
to the width of the transition features at the edges of the step in
central wavelength.

Hybrid differential stereoscopy reduces the effects by eliminating
the low spatial frequency components, so that the amplitude and slope
of the high spatial frequency component $\Lambda'_{H},$ rather than
of $\Lambda'$ itself, are of import: structures larger than roughly
the width of the patch used for correlation are attenuated or eliminated
by the high-pass filtering that generates $\Lambda'_{H}$.

\begin{figure}
\includegraphics[width=3in]{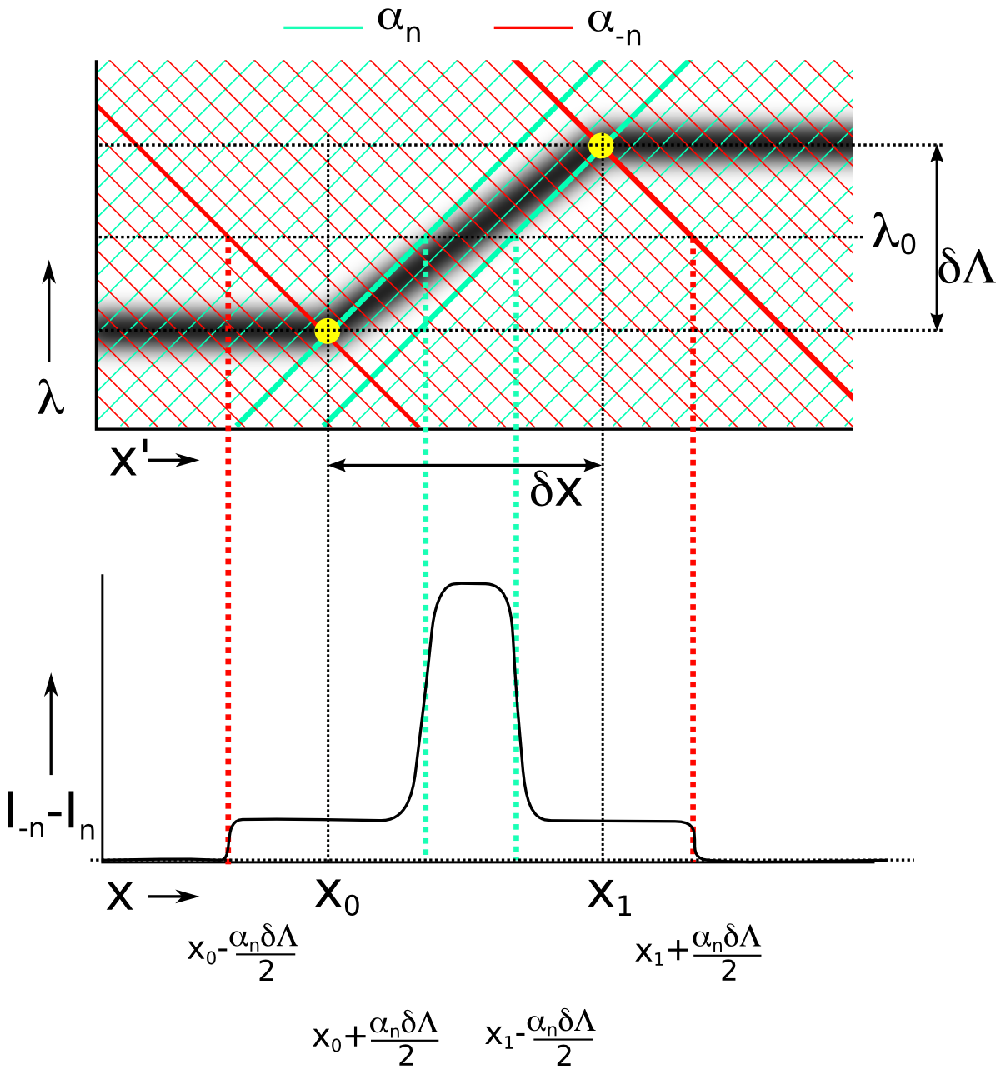}

\caption{\label{fig:step}A finite linear feature in $\Lambda_{0}$ demonstrates
the interplay between projection, brightness, and feature size in
a stereoscopic instrument. At top, a spectral line undergoes a finite
shift $\delta\Lambda$ between the points $x'_{0}$ and $x'_{1}$
in the ($x',\lambda$) plane. The instrument collects two spectral
orders with dispersion $\alpha_{\pm n}$. Each order integrates pixels
along lines of the appropriate slope: positive (cyan) or negative
(red) and projects onto the detector $x$ coordinate. At bottom the
line depth brightness integral $-(I_{n}-I_{-n})$ varies as a function
of pixel location (projected back to the $\lambda=\Lambda_{0}$ line
at top). }
\end{figure}

By considering carefully the geometry of the differential inversion
in the $(x,\lambda)$ plane it is possible to estimate the systematic
errors due to slope and line-displacement effects alone on differential
stereoscopy, even in the absence of image effects such as variations
in line or continuum intensity. Figure \ref{fig:step} shows the
geometry of dispersed integration through a spectral line. The contribution
$L'_{n}$ to the measured intensity $I_{n}$ is approximated to first
order in Eq. \ref{eq:dlambda/dx}, but the contribution may also be
written exactly in the case of constant $L(x)$, constant $d\Lambda'/dx$,
and a broad filter:\begin{equation}
I_{n}(x)=\int d\lambda\left(F(\lambda')E_{c}(x-\alpha_{n}\lambda')-L\delta(\lambda-\Lambda'(x)+\left(\alpha_{n}\lambda\right)d\Lambda'/dx\right)=C-\frac{L}{1+\alpha_{n}\left(d\Lambda'/dx\right)}\label{eq:analytic}\end{equation}
Applying this expression to Eq. \ref{eq:J-definition} gives an analytic
formula for the relation between J and $d\Lambda'/dx$. Letting $\gamma\equiv\alpha_{n}d\Lambda'/dx$
and $\ell\equiv L/C$, \begin{equation}
J_{n}=\frac{I_{n}-I_{-n}}{I_{n}+I_{n}}=-\gamma\left(\frac{\ell}{1-\ell}\right)\left(\frac{1}{1-\left(\gamma^{2}/(1-\ell)\right)}\right)\label{eq:j-analytic}\end{equation}
where the factor $\alpha_{n}\ell/(1-\ell)$ is a linear calibration
coefficient on $d\Lambda'/dx$ and the right-hand term is a nonlinearity
term. The nonlinearity in $J_{n}$ becomes strongly evident when $\gamma$
grows to about $0.5$, where (with $L/C=0.1$) the nonlinearity term
has a value of 1.38, and grows to infinity where $\gamma=\sqrt{(1-\ell)}$
. 

The strong nonlinearity in $J$ is partially compensated by spatial
distortion. The nonlinear term causes growth in $J$ compared to the
proportionality with $d\Lambda'/dx$, but that growth is accompanied
by spatial distortions that partially cancel the nonlinearity. In
particular, the measurable quantity of interest is $\Lambda'$ rather
than $d\Lambda'/dx$. In a pure differential inversion, with $\gamma$
positive and non-negligible, the $J_{n}$ signal arises from a slight
weakening of the spectral line in the $-n$ spectral order and a strong
strengthening in the $+n$ order. However, the image in the $+n$
channel is strongly foreshortened while the image in the $-n$ channel
is weakly fore-lengthened: the total $x$-integrated intensity in
each spectral order remains the same under small to moderate (<1 pixel)
$\Lambda'$ perturbations in the image. 

Consider a finite shift in central wavelength $\Lambda'$ as illustrated
in Figure \ref{fig:step}. Accompanying the shift in brightness is
a spatial distortion that affects the two spectral orders differently.
In the positive order the line depth is more strongly expressed, because
it is concentrated into fewer pixels than might be expected from the
spatial extent of the feature. Likewise, in the negative order the
line depth is more weakly expressed, because it is spread across more
pixels than the usual spatial extent of the feature. The corresponding
variation of $J_{n}$ thus has three regimes in this simple system
(neglecting the line width):\begin{equation}
J_{n}=\left\{ \begin{array}{cc}
\left|x\right|>\frac{\delta x+\alpha_{n}\delta\Lambda}{2} & 0\\
\frac{\delta x-\alpha_{n}\delta\Lambda}{2}\geq\left|x\right|>\delta x-\frac{\alpha_{n}\delta\Lambda}{2} & J_{mid}=\frac{I_{0}-I_{-n,l}}{I_{0}+I_{-n,l}}\\
\frac{\delta x-\alpha_{n}\delta\Lambda}{2}>\left|x\right| & J_{ctr}=\frac{I_{n,l}-I_{-n,l}}{I_{n,l}+I_{-n,l}}\end{array}\right.\label{eq:finite}\end{equation}
where (from Eq. \ref{eq:analytic}) $I_{0}=C-L$, $I_{-n,l}=C-L/(1+\gamma)$,
and $I_{n,l}=C-L/(1-\gamma)$. Simple analysis shows that\begin{equation}
J_{mid}=\frac{-\gamma}{(1+\gamma)(2/\ell-1)-1}\label{eq:jmid}\end{equation}
and \begin{equation}
J_{ctr}=\frac{-\gamma}{(1-\gamma^{2})/\ell-1},\label{eq:jctr}\end{equation}
so that the measured wavelength shift may be computed directly:

\begin{eqnarray}
\delta\Lambda_{meas.} & \equiv & -\alpha_{n}^{-1}\frac{C}{L}\int J_{n}dx\label{eq:deltalambdaintegral}\\
 &  & =-\alpha_{n}^{-1}\ell^{-1}\left\{ 2\alpha_{n}\delta\Lambda\left(\frac{-\gamma\ell}{(2-\gamma)(1-\ell)-\gamma}\right)+\delta x(1-\gamma)\left(\frac{-\gamma\ell}{(1-\gamma^{2})-\ell}\right)\right\} \nonumber \\
 &  & =\left\{ \left(\frac{2\gamma\delta\Lambda}{(2-\gamma)(1-\ell)-\gamma}\right)+\frac{(\delta x\alpha_{n}^{-1})(1-\gamma)(\gamma)}{1-\ell-\gamma^{2}}\right\} \nonumber \\
 &  & =\delta\Lambda\left\{ \frac{1}{1-\ell}\right\} \left\{ \frac{\gamma}{1+\gamma(1+\ell/2)}+\frac{(1-\ell)(1-\gamma)}{1-\ell-\gamma^{2}}\right\} \label{eq:deltalambda-exact}\\
 &  & \approx\delta\Lambda\left\{ \frac{1}{1-\ell}\right\} \left\{ \left(1-\frac{\ell\gamma}{1-\ell}\right)\right\} \label{eq:deltalambda}\end{eqnarray}
where the last row expands the $\gamma$-dependent error factor to
first order in $\gamma$. Equation \ref{eq:deltalambda} shows that
the linear approximations used to derive \ref{eq:hybrid_inversion}
are quite good, at least for this simple geometry: the first nontrivial
term in the error factor is two orders of approximation smaller than
the extracted signal (taking both $\gamma$ and $\ell$ to be small,
as did DeForest et al. 2004). The high quality of the approximation
can be seen in Figure \ref{fig:multi}: the measured peak and feature-averaged
$\Delta\Lambda_{meas}$ in each feature remains close to the injected
$\Delta\Lambda$ of the model, despite gross violation of the assumptions
used to derive Equation \ref{eq:hybrid_inversion}.

The astute reader will have noticed that Equation \ref{eq:deltalambda-exact}
has a pole at $\gamma=\sqrt{1-\ell}$. That is because we have neglected
finite length effects in the region where the line shift occurs. Equation
\ref{eq:deltalambda-exact} is more complex but exact for this geometry,
provided that the line width is small compared to $\delta x/\alpha_{n}$.
When that condition is violated (which happens when the center locus
is exceptionally narrow on the detector), individual pixel lines of
integration do not integrate across the entire foreshortened spectral
line but are limited instead by the ends of the spectral feature This
limitation keeps the signal response $J_{ctr}$ finite. Equation \ref{eq:deltalambda-exact}
neglects this effect, leading to a spurious pole. 

For nonzero line widths, Equation \ref{eq:deltalambda-exact} reveals
a constant correction factor of $1/(1-\ell)$ that must be applied
to stereoscopic inversions in general. This correction (of order 10\%)
is a true calibration coefficient, in the sense that it is independent
of $\gamma$.

\section{\label{sec:Summary-&-Specific}Summary \& Specific Observing Scenario}

The effects in \S\ref{sec:Hybrid-differential-stereoscopy} above
separate into two primary classes: additive and multiplicative noise
in the data. The additive effects, such as photon noise, affect sensitivity
for detection of small magnetic or Doppler features; the multiplicative
effects, affect the precision of magnetic measurements.  Here we calculate
a noise budget for a plausible observing scenario using a stereoscopic
instrument as a photospheric magnetograph. 

Consider an instrument observing quiet sun through the Dunn Solar
Telescope (DST; \citealt{Dunn1969}) and the medium-order adaptive
optics system (\citealt{Rimmele2004}), at the Fe I 617.34 nm line,
with 0.07 arcsec pixels (the diffraction limit is 0.2 arcsec), a 67
pm prefilter passband width detuned from $\Lambda_{0}$ by 6.7 pm
(10\% of the bandwidth), $\ell=0.1$, dispersion $\alpha=100$~pixel~nm$^{-1}$,
$n_{ph}=5\times10^{4}$ photons per pixel, and a correlation patch
FWHM of 15 pixels ($\epsilon=0.09$). Consider that a {}``typical''
resolved feature strength is between 100G-1kG, with length scales
of a few pixels (so that only stereoscopic measurements apply, without
the stabilizing influence of the more precise correlation stereoscopy).
The 617.34 nm line has a Land\'e $g$ factor of 2.5, so that the
Zeeman splitting equation is just '\begin{equation}
B'=\frac{hc}{4\Lambda_{0}^{2}g\mu}(\delta'\Lambda)=5.6T\, nm^{-1}\delta'\Lambda=5.6\, kG\,\AA^{-1}\delta'\Lambda\label{eq:splitting}\end{equation}
where $B'$ \textbf{}is measured resolution-smoothed magnetic field
strength ({}``flux density''), $h$ is Planck's constant, $c$ is
the speed of light, $g$ is the Land\'e factor, $\mu$ is the Bohr
magneton, and $\delta'\Lambda$ is the \emph{splitting} of the line
(twice the shift). The splitting measurement is the result of \emph{two}
independent measurements of a shift $\Delta\Lambda$: one in right-circularly
polarized light (Stokes I+V) and one in left-circularly polarized
light (Stokes I-V). Hence the noise $\Delta B'$ is a factor of $\sqrt{2}$
higher than might be supposed by simple scaling of the $\Delta\Lambda$
noise terms in \S\ref{sec:Hybrid-differential-stereoscopy} with
Equation \ref{eq:splitting}. 

The noise budget for a quiet-Sun DST observation may be summarized
as in -Table \ref{tab:additive-budget}. Two principal types of error
enter into the measurement: random or quasi-random noise as described
in \S\S\ref{sub:Photon-counting-noise}-\ref{sub:Image-brightness-gradients};
and calibration nonlinearities that enter as a result of filter detuning
and the first-order stereoscopic inversion as described in \S\S\ref{sub:Filter-passband-gradients}-\ref{sub:Rapid-variation-of}. 

\begin{table}
\begin{tabular}{|c|c|c|c|c|}
\hline 
Source&
Eq./Section&
$\Delta\Lambda$RMS (pm)&
$\Delta B'$RMS (G)&
cal. err. \tabularnewline
\hline
\hline 
Photon noise (diff.)&
\ref{eq:ph-noise}/\S\ref{sub:Photon-counting-noise}&
0.53&
$4.2$&
-\tabularnewline
\hline 
Photon noise (corr.)&
\ref{eq:correlation-photon-noise}/\S\ref{sub:Misalignment-noise}&
0.024&
$0.2$&
-\tabularnewline
\hline 
Image noise (corr.)&
\ref{eq:weighting-covariance}/\S\ref{sub:Image-noise-in}&
$0.12$&
1.0&
-\tabularnewline
\hline 
Continuum leakage&
\ref{eq:deltaj}/\S\ref{sub:Image-brightness-gradients}&
$0.25$&
$2.0$&
-\tabularnewline
\hline 
Line leakage &
\ref{eq:deltaj}/\S\ref{sub:Image-brightness-gradients}&
$0.15$&
1.2&
-\tabularnewline
\hline 
Filter signal leakage&
\ref{eq:asymmetric-diff}/\S\ref{sub:Filter-passband-gradients}&
-&
-&
3\%-7\%\tabularnewline
\hline 
Stereo nonlinearity&
\ref{eq:deltalambda}/\S\ref{sub:Rapid-variation-of}&
-&
-&
1\%-10\%\tabularnewline
\hline
\hline 
TOTAL (in quadrature)&
-&
$0.62$&
4.9&
3\%-12\%\tabularnewline
\hline
\end{tabular}

\caption{\label{tab:additive-budget}Noise budget for a sample stereoscopic
magnetograph measurement as described in the text. This measurement
would have \textasciitilde{} 6 G RMS noise even in field-free regions
of the Sun, and 3\%-12\% calibration errors in the measured flux of
small magnetized regions of 100-1000 G, no more than a few pixels
across. }
\end{table}

Because of the multiple sources of error, calculation of error budgets
for actual observations must be carried out on a case-by-case basis;
however, we note that for magnetograph observations of the photosphere,
the dominant noise term is (as expected) photon noise for a plausible
instrument setup and a range of targets that includes most non-sunspot-related
targets on the Sun. The expected formal single-exposure magnetogram
sensitivity is \textasciitilde{}10 Gauss (twice the RMS background
noise level) and the expected magnetic flux calibration accuracy is
expected to be close to 10\% in most cases including plage. 

The error budget in Table \ref{tab:additive-budget} is broadly consistent
with test measurements collected by DeForest et al. (2004) at the
Dunn Solar Telescope. In those measurements, exposures of only $\sim8\times10^{3}$
photons were collected; 16 exposures were combined to create each
image for an effective exposure of $1.2\times10^{5}$ photons; and
an \emph{asymmetric} inversion (whose sensitivity was reduced by a
factor of $\sqrt{2}$ compared to the symmetric inversion discussed
here) was used. The resulting calculated photon noise level is thus
a factor of $\sqrt{24/5}$ higher than shown here, or 9.5 G (0.45
G for the correlation noise), and the non-photon noise sources are
scaled up a factor of $\sqrt{2}$ compared to the sample measurement
described here. This leads to a total \emph{a priori} noise level
estimate of 10 G for that measurement. Indeed, in a small quiet region
with no detected flux features, DeForest et al. found \emph{a posteriori}
RMS noise levels between 9-12 Gauss, in good agreement with the current
\emph{a priori} calculation.

\section{\label{sec:Conclusions}Conclusions}

We have calculated a detailed noise budget for hybrid differential
inversion of the line-shift signal from a stereoscopic spectrograph,
with specific application as a stereoscopic magnetograph observing
a specific absorption line (Fe I 617.34 nm) in the solar photosphere.
The noise budget is photon dominated and careful \emph{a priori} analysis
shows that it offers similar calibration performance to existing and
past filtergraph (SOHO/MDI, \citealt{Scherrer1995}) and Fourier tachometer
(GONG; \citealt{Leibacher1999}) instruments (\citealt{Jones2001}):
instrument response to field is within a few to ten percent of linear
over a wide range of pixel-averaged field strengths.

The principal difference between the stereoscopic technique and the
filtergraph or Fourier tachometer techniques is that the instrument
can operate much more rapidly, because all required photons are collected
simultaneously in a dual-polarized-beam instrument. This is very important
for ground based observations in which atmospheric seeing effects
severely limit the precision with which line shifts can be measured
in time-multiplexing instruments such as filtergraphs. 

Some instruments, such as ZIMPOL (\citealt{Povel1998}), have been
able to limit crosstalk of atmospheric effects by multiplexing extremely
rapidly; but such instruments are still subject to motion blur even
in the absence of severe crosstalk. A stereoscopic spectral imager
should be able to acquire a diffraction limited magnetogram with few-Gauss
sensitivity in a small fraction of a second at most major observing
facilities. We expect this rapidity of acquisition to become important
even in space-based observing as higher resolution instruments (e.g.
\emph{Hinode/SOT}; \citealt{Ichimoto2006}) release data to the scientific
community: the race between solar evolution and data acquisition (photon
counting) give rise to a tradeoff between sensitivity and spatial
resolution for any spectral imager, even above the atmosphere. 

Particularly for features that are spatially small, stereoscopic imagers
offer a surprising advantage: because the fundamental signal extracted
from the images is proportional to the \emph{spatial derivative} of
the line shift signal, such instruments are particularly sensitive
to small features, right down to the spatial resolution of the measurement.
Like Dolby$^{\textrm{}{TM\}}}$ noise reduction for audio recordings
(e.g. \citealt{Vaseghy2006}), spatial derivative measurement enhances
high spatial frequencies, boosting the high frequency signal relative
to the photon noise floor. Conventional spectral measurements sample
each location in the image plane independently, so that in a roughly
circular feature of linear size $L$ and area $A$ the uncertainty
due to photon counting in the feature's total magnetic flux scales
as $A^{-1/2}$ or $L^{-1}$. In a differential stereoscopic measurement,
the \emph{average slope} uncertainty, rather than the total flux uncertainty,
scales as $L^{-1}$ so the total flux uncertainty is independent of
feature size for features smaller than the correlation patches described
in Section \ref{sec:Summary-&-Specific}.

Stereoscopic spectral imaging is a special case of the more general
technique of \emph{spectral tomography} (e.g. \citealt{Wilson1997}),
and appears to offer significant advantages in the regime we have
described: observations of a narrow spectral line, where only one
or two moments of the line are desired to be measured. Spectral tomography
using regularized inverse or conjugate analysis techniques (e.g. \citealt{Claerbout2004})
could potentially be used to extract much more information about the
spectrum, using multiple orders or, as did Wilson et al., crossed
gratings; but as the number of additional channels grows, so do the
potential paths for signal contamination and crosstalk between different
aspects of the measured spectrum. We conclude that stereoscopic techniques
represent a sweet spot for imaged Doppler or magnetic measurements
of solar scenes, offering superior photon efficiency and comparable
performance to filtergraph measurements. This is already important
as ground-based spectral imagers are already photon-starved near the
diffraction limit in existing meter-class telescopes (such as the
Dunn Solar Telescope, \citealt{Dunn1969}, and the Swedish Vacuum
Tower, \citealt{Scharmer2003}) and will grow more so near the diffraction
limit of future telescopes such as the four-meter Advanced Technology
Solar Telescope (\citealt{Keller2002}).

The authors thank Jack Harvey for helpful discussion of the current
state of magnetograph cross-calibration, and Steve Tomczyk for
pointing out the neceessity of this paper.  This work was funded
internally by the Southwest Research Institute.

\bibliographystyle{plainnat}

\end{document}